\documentclass{article}
 \usepackage{amsmath,amssymb,graphicx,cite}
  \title {Multiple-scale analysis of discrete nonlinear partial difference equations:
  the reduction of the lattice potential KdV. }
\DeclareMathAlphabet{\mathpzc}{OT1}{pzc}{m}{it}

\author{ D. Levi\\
Dipartimento di Ingegneria Elettronica, \\
Universit\'a degli
 Studi Roma Tre and Sezione INFN, Roma Tre,\\
Via della Vasca Navale 84,
 00146 Roma, Italy\\}
 \date{\today}

 \def\be{\begin{equation}}
\def\ee{\end{equation}}
 \def\ba{\begin{array}}
 \def\ea{\end{array}}
 \def\bea{\begin{eqnarray}}
 \def\eea{\end{eqnarray}}
 \def\beas{\begin{subequations}}
 \def\eeas{\end{subequations}}
 \def\bean{\begin{eqnarray*}}
 \def\eean{\end{eqnarray*}}

\begin{document}
 \maketitle
\begin{abstract}
 We consider multiple lattices and functions defined
on  them. We introduce  slow varying conditions for functions defined on the lattice and  express the
variation of a function  in terms of an asymptotic
expansion with respect to the slow varying lattices.

We use  these results to  perform the multiple--scale reduction of
the  lattice potential Korteweg--de Vries equation.
 \end{abstract}

 \section{Introduction}
 The reductive perturbation method (or
multiple--scale analysis) \cite{taniuti} allows us to deduce a set of simplified equations
starting from a basic model
without loosing its main characteristic features. The method consists
essentially in an asymptotic
analysis of a perturbation series, based on the existence of different scales to cure secularity.

The success of the method relies mainly on the nice property of the resulting reduced
models, which are simple and often integrable. Simple here means actually simpler
than the starting equations and still providing useful information.  Integrable means
that they carry an infinite set of conserved quantities, have an infinite set of symmetries and of exact solutions. Finally, as emphasized in \cite{calogero}, 
 the reductive perturbation approach preserve integrability. Consequently this approach can be used to obtain new  integrable models from known ones.

The situation is quite different in the case of differential
equations on a lattice (for example, in the case of dynamical systems when one has a
continuous time and discrete space variables) for which a reliable reductive
perturbative method which would produce reduced discrete systems
up to our knowledge does not exist. Leon and Manna \cite{leon} and later
Levi and Heredero \cite{lh} proposed a set of tools which allow 
to perform multiple--scale analysis for a discrete evolution equation.
These tools rely on the definition of a large grid scale via the
comparison of the magnitude of  related difference operators and on the introduction of a slow varying condition for function defined on the lattice.
Their results, however, are not very promising as the
reduced models are neither simpler nor more integrable than the
original one. Starting from an integrable model, like the Toda
lattice \cite{toda}, the Leon and Manna reduction technique produce a
non-integrable differential difference equation of the discrete
Nonlinear Schr\"odinger type \cite{yamilov,sakovich}. Levi and
Heredero \cite{lh} started from the integrable differential--difference 
Nonlinear Schr\"odinger equation and got a nonintegrable system of
differential--difference equations of Kortewg--de Vries type.

We consider here  the case of completely discrete equations
defined on a two dimensional orthogonal lattice. We follow the approach introduced by Levi and Heredero \cite{lh}, extended to the case of multiple orthogonal lattices.  We try to keep all passages consistent with the continuous limit,
when the lattice spacings on the different grids go to zero.

In Section \ref{rescaling} we introduce, following  \cite{lh}, the multiple lattices, the
slow varying conditions and the asymptotic expansions of the functions'
variations while in Section \ref{lkdv} we apply the resulting
formulas to the case of the multiple--scale expansion of the lattice potential Korteweg -- de Vries equation (lpKdV) \cite{frank,hietarinta},
\bea \label{3.1}
(p-q+u_{n,m+1}-u_{n+1,m})(p+q -u_{n+1,m+1} + u_{n,m}) = p^2 - q^2.
\eea
 At the end, in Section
\ref{conclusions} we discuss  the results obtained
and present a list of open problems and remarks relevant also for the case
 of a differential--difference
equation  \cite{lh}.


\section{Multi-lattice structure and the variation of a function on them} \label{rescaling}
 \subsection{Rescaling on the lattice} \label{2.1}
\begin{figure}
\centering
 \includegraphics[width=4in]{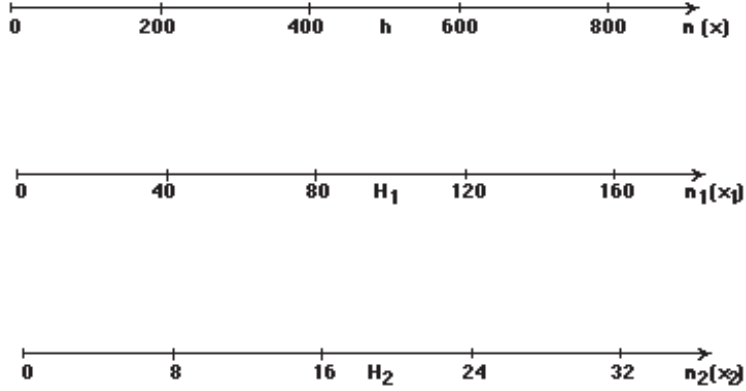}
\caption{Multiple lattices ($n,n_1=\frac{n}{N}, n_2=\frac{n}{N^2}$), with  $N=5$; the corresponding continuous coordinate would be $x = n h$, $x_1 = n_1 H_1$ and $x_2 = n_2 H_2$ where $H_1 = h N$, $H_2 = h N^2$.}
\label{fig1}
\end{figure}

Given a lattice defined by a constant lattice spacing $h$, we will
 introduce an apriory infinite number of lattices defined by 
 lattice spacings $H_j$, with $j=1,2,\ldots,\infty$, where $H_j$ are well defined functions of $h$, $H_j = H_j(h)$. In Fig. \ref{fig1} we show an example of such a situation with $j=1,2$.
 For convenience we will denote by $n_j$ the running index of the points separated by $H_j$
 and $n$ those separated by $h$. Moreover, in correspondence with the lattice variables,
 we can introduce the real variables $x = hn$ and
 $x_j=H_j n_j$.
 
A simple definition of $H_j$ is obtained by  introducing an integer number $N$ and defining
$H_j = N^j h$. If $N$ is a large number than $
\frac{1}{N} = \epsilon$ will be a small number.  The variables $x$ and $x_j$
will go over to continuous variables when respectively $h
\rightarrow 0$, $n \rightarrow \infty$ and  $H_j \rightarrow 0$,
$n_j \rightarrow \infty$ in such a way that their products $x=nh$
and $x_j=n_jH_j$ are finite.

Let us assume that  $x_j = \epsilon^j x$.
Then, if $x \sim \frac{1}{\epsilon^j}$,    $x_j \sim 1$.
 So $x_j$ represents, as $j$ increases an always larger
 portion of the $x$ axis.
    This assumption, together with the choice $
\epsilon=\frac{1}{N}$, will reflect onto a relation between the lattice variables $n$ and $n_j$ as
 \be \label{f2b}
 x_j = H_j n_j = h N^j n_j = \epsilon^j x = \frac{1}{N^j} h n \Rightarrow n_j ={\Bigl [} \frac{n}{N^{2j}} {\Bigr ]}.
 \ee
Consequently  we need to move $N^{2j}$ points on the lattice of the discrete variable   $n$ to shift the
lattice variable $n_j$ by $1$ point.
 
\subsection{Slowly varying functions and their expansion} \label{2.2}
Here we study the relation between functions acting on the different lattices defined in  Section \ref{2.1}.

Let us consider a function $f$ defined on the  points of a lattice of index $n$ given in Fig. \ref{fig1},
i.e. $f(n)$.
We are interested in understanding what happens when we assume
that $f(n) = g(n_1,n_2, \dots,n_K)$, i.e. $f$ depends on a finite
number $K$ of slow varying lattice variables $n_j$  such that $g(n_j \pm k) = f(n \pm k N^{2j})$ (\ref{f2b}). As we are mainly interested in applying  in Section \ref{lkdv} these results to the lpKdV  (\ref{3.1}), we need to know what happens to the function $g$ when the function
$f$ is in the point $n + 1$. One needs to get explicit expressions for $f(n+1)$ in terms of $g(n_1, n_2, \ldots, n_K)$ on different points in the $n_1$, $n_2$, $\ldots, n_k$ lattices. At first let us study the case, 
considered in \cite{lh} when we have only two different lattices,
i.e. $K = 1$. Using the results obtained in this case we will then
consider the case corresponding to $K = 2$. The generic case will than be obvious.

In the case of one variable we can use the result contained in
\cite{Jordan}:
\begin{equation}\label{difInt}
\Delta_H^k g(n_1)=\sum_{i=k}^\infty  {k!\over i!}P(i,k)\Delta^i_h
f(n)
\end{equation}
where $H$ is any one of the possible $H_j$ introduced before and
 $\Delta_H^k g(n_1) = \sum_{i=0}^{k}  (-1)^{k-i}
{k \choose i} g(n_1+i)$, the $k$--variation formula obtained
using a two--points forward difference scheme. The coefficients
$P(i,k)$ are given by
\begin{gather*}
P(i,j)=\sum_{\alpha=j}^i
\left(\frac{H}{h}\right)^\alpha S_{i}^{\alpha}\mathfrak{S}_{\alpha}^j
\end{gather*}
with~$S_{i}^{j}$, $\mathfrak{S}_{i}^j$ Stirling numbers of the
first and second kind respectively. A table with the coefficients
$P(i,k)$ for $(i,k) < (6,6)$ is contained in ref \cite{Jordan}.

Eq. (\ref{difInt}) allow us to express a difference of
order $k$ in the lattice of spacing $H$ in terms of an infinite
number of differences on the lattice of spacing $h$.
To get an approximate solution we have to truncate the expansion at the r.h.s. of eq. (\ref{difInt})
by requiring a slow varying condition for the function $f(n)$. {\it We
will say that the function $f(n)$ is a slow varying function of
order $p$ if $\Delta^{p+1}_h f(n)=0.$} A slow varying function of
order $p$ is a polynomial of degree $p$ in $n$ \cite{tit}. For a function of order ~$p=1$, eq. (\ref{difInt}) reduces to
\bea \label{2.aa}
 \Delta_H g(n_1)=N^2 \Delta_h f(n).
 \eea
Dividing eq. (\ref{2.aa}) by $h$ and taking the limit as $h \rightarrow 0$, with $x=hn$ and $x_1 = n_1 H_1 = n_1 h N$ finite, we get $\frac{df(x)}{dx} = \epsilon \frac{dg(x_1)}{dx_1}$. In the case~$p=2$, we  get:
\bea \label{2a} 
&& \Delta^2_H g(n_1)=N^4 \Delta^2_h f(n),\\
\label{2b} &&\Delta_H g(n_1)= N^2 \Delta_h f(n)+\frac{N^2 (N^2
-1)}{2}\Delta^2_h f(n). 
\eea
 For~$p=3$ we have
\begin{gather*}
 \Delta^3_H g(n_1)=N^6 \Delta^3_h f(n),\\
 \Delta^2_H g(n_1)=N^4 \Delta^2_h f(n)+ N^4 (N^2 - 1)
\Delta^3_h f(n),\\
 \Delta_H
g(n_1)= N^2 \Delta_h f(n)+\frac{N^2 (N^2 - 1)}{2}\Delta^2_h f(n)
+\frac{N^2 (N^2 - 1) (N^2 - 2)}{6}\Delta^3_h f(n),
\end{gather*}
and for $p=4$ we have:
\bea \label{p4}
 &&\Delta^4_H g(n_1)=N^8 \Delta^4_h f(n),\\ \nonumber
 &&\Delta^3_H g(n_1)=N^6 \Delta^2_h f(n)+\frac{3 N^6 (N^2 - 1)}{2}
\Delta^4_h f(n),\\ \nonumber && \Delta_H^2
g(n_1)= N^4 \Delta_h^2 f(n)+ N^4 (N^2 - 1)\Delta^3_h f(n)+\\
\nonumber && \qquad +[\frac{7}{12} N^6 (N^2 - 1) -
\frac{11}{12} N^4 (N^2 - 1)]\Delta^4_h f(n),\\
\nonumber && \Delta_H g(n_1) = N^2 \Delta_h f(n) + \frac{1}{2!} N^2
(N^2 - 1)\Delta^2_h
f(n) + \\
\nonumber && +\frac{1}{3!}N^2 (N^2 - 1) (N^2 - 2) \Delta^3_h f(n)+
\frac{1}{4!}N^2 (N^2 - 1) (N^2 - 2) (N^2 - 3) \Delta^4_h f(n).
\eea

 From (\ref{2.aa}),   if 
$f(n)$ is a slow function of order 1, $f(n+1)$ reads:
\be \label{f5}
f(n+1) = g(n_1) + \frac{1}{N^2} [g(n_1+1)- g(n_1)]
+ o(\frac{1}{N^4}),
\ee
while, if the function $f(n)$ is a slow varying function of order 2, $f(n+1)$
is given by
\bea \label{f6}
f(n+1) &=& g(n_1)
+\frac{1}{2 N^2} [ - g(n_1+2) + 4 g(n_1+1) -3 g(n_1)] +\\
\nonumber &+& \frac{1}{2 N^4} [ g(n_1+2) -2 g(n_1+1) + g(n_1)] +
o(\frac{1}{N^6}).
\eea
When the function $f(n$) is a slow varying function of order 3,  $f(n+1)$ is
given by
\bea \nonumber
f(n+1)& =& g(n_1)
+\frac{1}{6 N^2} [ 2 g(n_1+3) - 9 g(n_1+2) + 13 g(n_1+1) -6 g(n_1)] +\\
 \label{f7a} &+&
\frac{1}{2 N^4} [ - g(n_1+3)+ 4 g(n_1+2) - 5 g(n_1+1) + 2 g(n_1)] +\\
\nonumber  &+& \frac{1}{6 N^6} [ g(n_1+3)  -3 g(n_1+2) +3 g(n_1+1)
- g(n_1)] + o(\frac{1}{N^8}),
\eea
and, when the function $f(n)$ is a slow varying function of order 4, 
is given
\bea\label{f7} &f(n+1) = g(n_1)+& \\ \nonumber
&+\frac{1}{12 N^2} [ - 3 g(n_1+4) + 16 g(n_1+3) - 36 g(n_1+2) + 48 g(n_1+1) - 25 g(n_1)] +&\\
\nonumber &+
\frac{1}{24 N^4} [ 11 g(n_1+4) - 56 g(n_1+3)+ 114 g(n_1+2) - 104 g(n_1+1) + 35 g(n_1)] +&\\
 \nonumber &+ \frac{1}{12 N^6} [ -3 g(n_1+4) +14 g(n_1+3)  - 24
g(n_1+2) + 18 g(n_1+1)
- 5 g(n_1)] +&\\
\nonumber  &+ \frac{1}{24 N^8} [ g(n_1+4) -  4 g(n_1+3)+ 6 g(n_1+2)
- 4 g(n_1+1) +  g(n_1)] + o(\frac{1}{N^{10}}).&
 \eea

In  Section \ref{lkdv} we consider the reduction of an integrable discrete equation and will be interested in obtaining from it  integrable discrete equations. It is known  \cite{ly} that a scalar differential difference equation can possess higher conservation laws and thus be integrable only if
it depends symmetrically on the discrete variable, i.e. if the discrete equation is invariant with respect to the
inversion of $n$. So  we will choose asymptotic discrete formulas which contain both $f(n \pm 1)$. The results
contained in eq. (\ref{difInt}) do not provide us with centralized
formulas. To get symmetric formulas we  need to take into account the
following observations:
\begin{enumerate}
    \item Eq. (\ref{difInt}) is valid also if $H$ and $h$ are both
    negative;
    \item For a slow varying function of order $p$, $\Delta_h^p
    f(n) = \Delta_h^p
    f(n+l)$ for any integer number $l$.
\end{enumerate}
Using these observations, from eq. (\ref{f5}) we get
\be \label{f5a}
f(n-1) = g(n_1) + \frac{1}{N^2} [g(n_1-1)- g(n_1)]
+ o(\frac{1}{N^4}),
\ee and
in place of eq. (\ref{f6}) we
have
\bea \label{f10}
f(n+1) &=& g(n_1) +  \frac{1}{2 N^2}[g(n_1+1) - g(n_1-1)] + \\ \nonumber \qquad &+& \frac{1}{2 N^4}[g(n_1+1) - 2 g(n_1)+ g(n_1-1)] +
o(\frac{1}{N^6}),
\eea
when the function $f(n)$ is a slow varying function of order 2. To
 get eq.
(\ref{f10}) we have to write, using Observation 2,   eqs. (\ref{2a}, \ref{2b})
in the form
\bea \label{10} 
&&\Delta_H^2 g(n_1)= N^4 {\tilde \Delta}^2_h f(n),
\\ \label{10a} &&\Delta_H g(n_1)= N^2 \Delta_h f(n)+\frac{N^2 (N^2
-1)}{2}{\tilde \Delta}^2_h f(n),
\eea
where by the symbol $\tilde \Delta$ we mean the centralized
version of the difference operator. In the case of eq. (\ref{10}, \ref{10a}) ${\tilde \Delta}^2_h f(n) = f(n+1) - 2 f(n) + f(n-1)$.

When  $f(n)$ is a slow varying function of odd order  we are not able
to construct completely symmetric  derivatives using
just two points forward difference formulas and thus $f(n+1)$ and
$f(n-1)$ will never be expressed in a symmetric form (see for example eqs. (\ref{f5}, \ref{f5a})). In the case
when the function $f(n)$ is a slow varying function of order 4, we
have to rewrite formulas (\ref{p4}) using both observations
introduced before. We have:
\bea \label{p10} && {\tilde \Delta}^4_H g(n_1)=N^8 {\tilde
\Delta}^4_h f(n),\\ \nonumber &&{\tilde \Delta}^2_H g(n_1)=N^4
{\tilde \Delta}^2_h f(n)+\frac{ N^4 (N^4 - 1)}{12} {\tilde
\Delta}^4_h f(n),\\ \nonumber
 &&\Delta_H^2
g(n_1)= N^4 \Delta_h^2 f(n)+ N^4 (N^2 - 1){\hat \Delta}^3_h f(n) +\\
\nonumber && \qquad +[\frac{7}{12} N^6 (N^2 - 1) -
\frac{11}{12} N^4 (N^2 - 1)] {\tilde \Delta}^4_h f(n),\\
\nonumber
 &&\Delta_H
g(n_1) = N^2 \Delta_h f(n)  +\frac{1}{3!}N^2 (N^2 - 1) (N^2 - 2) {\hat \Delta}^3_h f(n) +\\
\nonumber &&+ \frac{1}{2!} N^2 (N^2 - 1)\Delta^2_h
f(n) + \frac{1}{4!}N^2 (N^2 - 1) (N^2 - 2) (N^2 - 3) {\tilde
\Delta}^4_h f(n). \eea
where, using Observation 2, we have written $\Delta^3_h f(n)$ as
${\hat \Delta}^3_h f(n) = 2 f(n+2) -7 f(n+1) + 9 f(n) - 5 f(n-1) +
f(n-2)$. Inverting formulas (\ref{p10}) we have:
\bea \label{f14}
f(n+1) = g(n_1)   - \frac{1}{12 N^2} [ g(n_1+2) - 8 g(n_1+1) + 8
g(n_1-1) - g(n_1-2) ]
\\
\nonumber - \frac{1}{24 N^4} [ g(n_1+2) - 16 g(n_1+1) + 30 g(n_1)
- 16 g(n_1-1) +
g(n_1-2) ] \\
\nonumber  + \frac{1}{12 N^6} [ g(n_1+2) - 2 g(n_1+1) + 2 g(n_1-1) - g(n_1-2) ]+  \\
\nonumber   \frac{1}{24 N^8} [ g(n_1+2) - 4 g(n_1+1) + 6 g(n_1) -
4 g(n_1-1) + g(n_1-2) ] + o(\frac{1}{N^{10}}).
\eea

Let us pass now to the case of functions of multiple variables,
i.e.  when  $g=g(n_1,n_2, \ldots, n_K)$. If $n_1$,
$n_2$, \ldots , $n_K$ were completely independent discrete
variables than any operation on one of the variables will not
reflect on the other. If, however  we have $f(n) =
g(n_1,n_2, \ldots, n_K)$ and $n_j = \frac{n}{N^{2j}}$, then any shift of $n$ will reflect on
all variables $n_1$, $n_2$, \ldots , $n_K$. We will consider later the case when the multiple variables are independent, i.e. the partial difference case.

Let us consider here in all details the case
of $K=2$, which is the case we will need in the Section \ref{lkdv}.  $f(n) = g(n_1,n_2)$ and we are looking for a representation of $f(n+1)$ in terms of $g(n_1,n_2)$ and its shifted values. The resulting formulas will depend in a crucial way on the slow varying order of the function $f(n)$ with respect to $n_1$ and $n_2$. If the variation of both variables has to appear in the expansion of  $f(n + 1)$ than  the slow varying order with respect to $n_1$ must be greater than that of $n_2$. $f(n)$ cannot be a slow varying function of order 1 in $n_1$ as in this case $f(n + 1)$ will have no variation in $n_2$. If  $f(n)$ is a slow varying function of order 2 in $n_1$ than it can be either of order 1 or 2 in $n_2$. In both cases the obtained formula will be valid up to order $\frac{1}{N^4}$, but, in the first case the obtained expression will not be symmetric in $n_2$.  When $f(n)$ is a slow varying function of order 2 in $n_1$ and of order 1 in $n_2$, taking into account eqs. (\ref{10a}, \ref{2.aa}) and the observation given before, we have:
\beas \label{h15}
\begin{gather} \label{h15a}
g(n_1+1,n_2) = g(n_1,n_2) + N^2 [f(n+1,n_2)   - f(n,n_2)] + \\
\nonumber + \frac{N^2 ( N^2 -1)}{2}[ f(n+1,n_2) - 2 f(n,n_2) +
f(n-1,n_2)],
\\ \label{h15b}
g(n_1-1,n_2) = g(n_1,n_2) - N^2 [f(n+1,n_2)   - f(n,n_2)] + \\
\nonumber + \frac{N^2 ( N^2 -1)}{2}[ f(n+1,n_2) - 2 f(n,n_2) +
f(n-1,n_2)],
\\ \label{h15c}
g(n_1,n_2+1) = g(n_1,n_2) + N^4 [f(n_1,n+1)   - f(n_1,n)],
\\ \label{h15e}
g(n_1+1,n_2+1) = g(n_1,n_2+1) + N^2 [f(n+1,n_2+1)   -
f(n_1,n_2+1)] + \\
\nonumber +\frac{N^2 ( N^2 -1)}{2}[ f(n+1,n_2+1) - 2 f(n_1,n_2+1)
+ f(n-1,n_2+1)],
\\ \label{h15f}
g(n_1-1,n_2+1) = g(n_1,n_2+1) - N^2 [f(n+1,n_2+1)   -
f(n_1,n_2+1)] + \\
\nonumber +\frac{N^2 ( N^2 -1)}{2}[ f(n+1,n_2+1) - 2 f(n_1,n_2+1)
+ f(n-1,n_2+1)],
\end{gather}
\eeas
where we took  into account that on the unshifted point $n
= n_1 = n_2$, $f(n)=f(n,n_2)=f(n_1,n)=f(n,n)=g(n_1,n_2)$ and that $f(n + 1, n_2 + 1)$, which is appearing in eqs.
(\ref{h15e}, \ref{h15f}), is obtained from $g(n_1, n_2 + 1)$,
given by eq. (\ref{h15c}),  by substituting $n_1$ by $n+1$.
 In such a way the 5 variables on
the l.h.s. of eqs. (\ref{h15}) are expressed in terms of the 6
variables  $f(n)$, $f(n + 1)=f(n + 1, n + 1)$,
$f(n_1,n + 1)=f(n,n + 1)$, $f(n \pm 1, n_2)=f(n \pm 1, n) $, $f(n-1,n+1)$ 
appearing on the r.h.s.  of eqs. (\ref{h15}).  Let us notice that to get a coherent number of equations with respect to the unknowns we had to consider also eqs. (\ref{h15b}, \ref{h15f}) which involve $n_1-1$. We can invert the system (\ref{h15}) and get:
\bea \label{h16}
f(n+1) = g(n_1,n_2) + \frac{1}{ 2 N^2} [g(n_1+1,n_2)   - g(n_1-1,n_2)] + \\ \nonumber 
+ \frac{1}{ N^4} 
[g(n_1,n_2+1) - g(n_1,n_2) ]  +\\
\nonumber + \frac{1}{2 N^4}\{ [g(n_1-1,n_2) - 2 g(n_1,n_2) +
g(n_1+1,n_2)]\}+ 
o(\frac{1}{N^{6}}).
\eea

In the continuous limit, when $n \rightarrow \infty$ and $h
\rightarrow 0$ in such a way that $nh \rightarrow x$,  eq.
(\ref{h16}) will give
\be \label{f17} f_{,x}(x) = \epsilon g_{,x_1}(x_1,x_2) +
\epsilon^2 g_{,x_2}(x_1,x_2). \ee

When  $f(n)$ is a slow varying function of second order in both variables, in place of eq. (\ref{h15}) we have:
\beas \label{f15}
\begin{gather} \label{f15a}
g(n_1+1,n_2) = g(n_1,n_2) + N^2 [f(n+1,n_2)   - f(n,n_2)] + \\
\nonumber + \frac{N^2 ( N^2 -1)}{2}[ f(n+1,n_2) - 2 f(n,n_2) +
f(n-1,n_2)],
\\ \label{f15b}
g(n_1-1,n_2) = g(n_1,n_2) - N^2 [f(n+1,n_2)   - f(n,n_2)] + \\
\nonumber + \frac{N^2 ( N^2 -1)}{2}[ f(n+1,n_2) - 2 f(n,n_2) +
f(n-1,n_2)],
\\ \label{f15c}
g(n_1,n_2+1) = g(n_1,n_2) + N^4 [f(n_1,n+1)   - f(n_1,n)] + \\
\nonumber + \frac{N^4 ( N^4 -1)}{2}[ f(n_1,n+1) - 2 f(n_1,n) +
f(n_1,n-1)],
\\ \label{f15d}
g(n_1,n_2-1) = g(n_1,n_2) - N^4 [f(n_1,n+1)   - f(n_1,n)] + \\
\nonumber + \frac{N^4 ( N^4 -1)}{2}[ f(n_1,n+1) - 2 f(n_1,n) +
f(n_1,n-1)],
\\ \label{f15e}
g(n_1+1,n_2+1) = g(n_1,n_2+1) + N^2 [f(n+1,n_2+1)   -
f(n_1,n_2+1)] + \\
\nonumber +\frac{N^2 ( N^2 -1)}{2}[ f(n+1,n_2+1) - 2 f(n_1,n_2+1)
+ f(n-1,n_2+1)],
\\ \label{f15f}
g(n_1-1,n_2+1) = g(n_1,n_2+1) - N^2 [f(n+1,n_2+1)   -
f(n_1,n_2+1)] + \\
\nonumber +\frac{N^2 ( N^2 -1)}{2}[ f(n+1,n_2+1) - 2 f(n_1,n_2+1)
+ f(n-1,n_2+1)],
\\ \label{f15g}
g(n_1+1,n_2-1) = g(n_1,n_2-1) + N^2 [f(n+1,n_2-1)   -
f(n_1,n_2-1)] + \\
\nonumber +\frac{N^2 ( N^2 -1)}{2}[ f(n+1,n_2-1) - 2 f(n_1,n_2 -
1) + f(n-1,n_2-1)],
\\\label{f15h}
g(n_1-1,n_2-1) = g(n_1,n_2-1) - N^2 [f(n+1,n_2-1)   -
f(n_1,n_2-1)] + \\
\nonumber +\frac{N^2 ( N^2 -1)}{2}[ f(n+1,n_2-1) - 2 f(n_1,n_2 -
1) + f(n-1,n_2-1)].
\end{gather}
\eeas
 In this case the 8 variables on
the l.h.s. of eqs. (\ref{f15}) are expressed in terms of the 9
variables  $f(n)=f(n, n)$, $f(n \pm 1)=f(n \pm 1, n \pm 1 )$,
$f(n_1,n \pm 1)=f(n,n \pm 1)$, $f(n \pm 1, n_2)=f(n \pm 1, n)$, $f(n-1,n+1)$ and $f(n+1,n-1)$
appearing on the r.h.s. of eqs. (\ref{f15}). One can invert the system (\ref{f15})  and
gets:
\bea \label{f16}
f(n+1) &=& g(n_1,n_2) + \frac{1}{ 2 N^2} [g(n_1+1,n_2)   - g(n_1-1,n_2)] + \\
\nonumber &+& \frac{1}{2 N^4} [g(n_1-1,n_2) - 2 g(n_1,n_2) +
g(n_1+1,n_2)] +\\ \nonumber &+& \frac{1}{2 N^4}[g(n_1,n_2+1) - g(n_1,n_2-1) ]+ 
o(\frac{1}{N^{6}}).
\eea

In the continuous limit, when $n \rightarrow \infty$ and $h
\rightarrow 0$ in such a way that $nh \rightarrow x$, eq.
(\ref{f16}) will give eq. (\ref{f17}).

Let us write here just the final result when
 $f(n)=g(n_1,n_3)$ is of order 4 in the
variable $n_1$ and of order 2 in $n_3$. We have:
\bea \label{f18}
&&f(n+1) = g(n_1,n_3) -\\ \nonumber
&& - \frac{1}{ 12 N^2} [g(n_1+2,n_3) - 8 g(n_1+1,n_3) + 8 g(n_1-1,n_3)   - g(n_1-2,n_3)] + \\
\nonumber &&+ \frac{1}{24 N^4} [g(n_1+2,n_3) - 16 g(n_1 + 1,n_3) + 30 g(n_1,n_3) - 16
g(n_1-1,n_3) + \\ \nonumber &&+ g(n_1-2,n_3)] +  \frac{1}{2 N^6} [ g(n_1,n_3+1) - g(n_1,n_3-1) ] - \frac{1}{12 N^6} [ g(n_1+2,n_3)   \\ \nonumber && - 2 g(n_1+1,n_3) + 2 g(n_1-1,n_3) - g(n_1-2,n_3) ] + 
o(\frac{1}{N^{8}}).
\eea
In the continuous limit, when when $n \rightarrow \infty$ and $h
\rightarrow 0$ in such a way that $nh \rightarrow x$,  eq.
(\ref{f18}) will give
\be \label{f19} f_{,x}(x) = \epsilon g_{,x_1}(x_1,x_3) +
\epsilon^3 g_{,x_3}(x_1,x_3). \ee

One can introduce  constant parameters in the definition of $n_1$, $n_2$  or $n_3$ in terms of $n$. For example we can write $n_1 = \frac{n M_1}{N^2}$ and $n_2 = \frac{n M_2}{N^4}$. $M_1$ and $M_2$ cannot be completely arbitrary as $n_1$, $n_2$, $n$ and $N$ are integers. In such a case eq. (\ref{h16}) reads:
\bea \label{h16a}
f(n+1) &=& g(n_1,n_2) + \frac{M_1}{ 2 N^2} [g(n_1+1,n_2)   - g(n_1-1,n_2)] + \\ \nonumber 
&+& \frac{M_2}{ N^4} 
[g(n_1,n_2+1) - g(n_1,n_2) ]  +\\
\nonumber &+& \frac{M_1^2}{2 N^4}\{ [g(n_1-1,n_2) - 2 g(n_1,n_2) +
g(n_1+1,n_2)]\}+ 
o(\frac{1}{N^{6}}).
\eea

In  Section \ref{lkdv} we will apply these results to a partial difference equation. For the sake of simplicity from now on we write the independent variables as indices. For completeness, in the following we present the formulas for  two independent lattices, $n$ and $m$,  and a function defined on them $f_{n,m}$. As the two lattices are independent the formulas presented above apply independently on each of the lattice variables. So, for example, $f_{n+1,m}$ when the function $f$ is a slowly varying function of order 2 of a lattice variable $n_1$ (see eq. (\ref{f10}) ) will read:
\bea \label{f100}
f_{n+1,m} &=& g_{n_1,m} +  \frac{1}{2 N^2} [g_{n_1+1,m} - g_{n_1-1,m}] + \\
\nonumber &+& \frac{1}{2 N^4}[g_{n_1+1,m} - 2 g_{n_1,m}+ g_{n_1-1,m}] +
o(\frac{1}{N^6}),
\eea
and similarly for a variation with respect to $m$ alone or to the case when we will introduce multiple lattices associated to $n$ or $m$, when formulas (\ref{h16}, \ref{f16}, \ref{f18}) are to be taken into account. A slightly less obvious situation appears when we consider $f_{n+1,m+1}$, as new terms will appear. We consider here just the case we will need later when $n_1 = \frac{M_1 n}{N^2}$, $m_1 = \frac{M_2 m}{N^2}$ and $m_2 = \frac{n}{N^4}$. If $f$ is a slow varying function of first order in $m_2$ and of second order in both $n_1$ and $m_1$,  $f_{n+1,m+1}$ reads:
\bea \label{h16b}
f_{n+1,m+1} &= & g_{n_1,m_1,m_2} + \frac{M_1}{ 2 N^2} [g_{n_1+1,m_1,m_2}   - g_{n_1-1,m_1,m_2}] + \\ \nonumber 
&+& \frac{M_2}{ 2 N^2} [g_{n_1,m_1+1,m_2}   - g_{n_1,m_1-1,m_2}] + \\ \nonumber
&+& \frac{M_1^2}{ 2 N^4} [g_{n_1+1,m_1,m_2}   + g_{n_1-1,m_1,m_2} - 2 g_{n_1,m_1,m_2}] + \\ \nonumber 
&+& \frac{M_2^2}{ 2 N^4} [g_{n_1,m_1+1,m_2}   + g_{n_1,m_1-1,m_2} - 2 g_{n_1,m_1,m_2}] + \\ \nonumber 
&+& \frac{M_1 M_2}{ 4 N^4} [g_{n_1+1,m_1+1,m_2}   + g_{n_1-1,m_1-1,m_2} - g_{n_1+1,m_1-1,m_2} - \\ \nonumber &-& g_{n_1-1,m_1+1,m_2}] +  \frac{1}{  N^4}
[g_{n_1,m_1,m_2+1} - g_{n_1,m_1,m_2} ]  +o(\frac{1}{N^{6}}).
\eea
As one can see in its fifth  and sixth lines,  eq. (\ref{h16b}) contains  extra terms involving variations in both the $m_1$ and the $n_1$ lattices.
\section{Reduction of the lattice potential KdV} \label{lkdv}

In this Section we apply the results presented in Section 2 to the case of the lattice potential KdV (\ref{3.1}).
By expanding the left hand side of eq. (\ref{3.1}), one separates the linear and nonlinear parts:
\bea \label{3.2}
(p-q) (u_{n+1,m+1} - u_{n,m})+ (p+q) (u_{n+1,m}-u_{n,m+1})= \\ \nonumber (u_{n+1,m}-u_{n,m+1})(u_{n+1,m+1} - u_{n,m}).
\eea
This equation involves just four points which lay on two  orthogonal infinite lattices and are the vertices of an elementary square.

Let us solve the linear equation 
\bea \label{3.2a}
F = (p-q) (u_{n+1,m+1} - u_{n,m})+ (p+q) (u_{n+1,m}-u_{n,m+1})= 0.
 \eea
The discrete Fourier transform \cite{tit} will reduce the solution of the Partial Difference Equation (P$\Delta$E) (\ref{3.2a}) to that of an ordinary difference equation.
Defining:
\bea \label{3.3}
u_{n,m} &=& \frac{1}{2 \pi i} \oint_{{\cal C}_1} v_m(z) z^{n-1} dz, \\ \label{3.4}
v_m(z) &=& \sum_{n=-\infty}^{+\infty} u_{n,m} ~ z^{-n}, 
\eea
where ${\cal C}_1$ is the unit circle, we reduce equation (\ref{3.2a}) to the following first order equation for $v_m(z)$
\bea \label{3.5}
v_{m+1}(z) [ (p-q) z - (p+q)] - v_m(z) [(p-q) - z (p+q)] = 0, 
\eea
whose solution is given by
 \bea \label{3.6}
v_m(z) = {\Bigl [} \frac{(p-q) - z (p+q)}{(p-q) z - (p+q)}{\Bigr ]}^m v_0(z). 
\eea
Given any initial condition $u_{n,0}$ the general solution of (\ref{3.2a}) is given by
 \bea \label{3.7}
u_{n,m} = \frac{1}{2 \pi i} \sum_{j=-\infty}^{+\infty} u_{j,0} \oint_{{\cal C}_1} {\Bigl [} \frac{(p-q) - z (p+q)}{(p-q) z - (p+q)}{\Bigr ]}^m   ~ z^{n-j-1} dz. 
\eea
Eqs. (\ref{3.3},  \ref{3.7}) can be rewritten in a more natural way (from the continuous point of view) by defining 
\bea \label{3.8}
z = e^{ik}; \qquad  \Omega = e^{-i \omega} = {\Bigl [} \frac{(p-q) - z (p+q)}{(p-q) z - (p+q)}{\Bigr ]}.
\eea
 In such a case eq. (\ref{3.3}) is just the standard Fourier transform and the solution (\ref{3.7}) is just written as a superposition of linear waves. The dispersion relation for these linear waves is given by $\omega = \omega(k)$ and reads:
 \bea \label{3.9}
 \omega = -2 \arctan {\Bigl [} \frac{p}{q} \tan {\Bigl (} \frac{k}{2} {\Bigr )} {\Bigr ]}.
 \eea 
In the following, however, to avoid too complicate formulas we express the solutions of the linear equation in term of $z$ and $\Omega$.
 
The lpKdV is an integrable equation of the same category of the KdV \cite{cd} as it possesses a Lax pair \cite{frank1} which can be obtained by  requiring that the model be {\sl consistent around a cube}.  So, as from KdV we get by multiple-scale reduction the NLS \cite{kz}, the same we may expect here \cite{calogero}. To get an integrable discrete equation we expect  a resulting discrete equation which is  somehow symmetric. At least when  $h_t \rightarrow 0$ with $t = m h_t$ the differential difference equation we obtain must be symmetric in terms of the inversion of $n_j$ \cite{ly}, i.e. if it depends on $n_{j+k}$ it will depend in the same way on $n_{j-k}$. So in the transformation of the discrete dependent variables we prefere to use formulas (\ref{f10}, \ref{f14}) for the space variables, while considering the lowest possible approximation for the discrete time variable (\ref{f5}). So, in the multiple--scale expansion, as we do not need to have the discrete time variable appearing in a symmetric way, we use eqs. (\ref{h16}).

Taking into account eq. (\ref{3.7}), we consider a wave solution of  eq. ( \ref{3.2a}) given by
\bea \label{3.8a}
E_{n,m} = e^{i(kn-\omega(k) m)} = z^n \Omega^m.
\eea
 Eq. (\ref{3.8a}) solves $F=0$ if $\omega$ is given by eq. (\ref{3.9}).
 Then we look for solutions of eq. (\ref{3.2}) written as a combination of modulated waves:
\bea \label{3.9a}
u_{n,m} = \sum_{s=0}^{+\infty} \varepsilon^{\beta_s} \psi^{(s)}_{n_1,m_1,m_2} (E_{n,m})^s +\\ \nonumber + \sum_{s=0}^{+\infty} \varepsilon^{\beta_s} {\bar \psi}^{(s)}_{n_1,m_1,m_2} ({\bar E}_{n,m})^s 
\eea
where $\psi^{(s)}_{n_1,m_1,m_2}$ are slowly varying functions on the lattice and $\varepsilon = N^{-2}$. By  $\bar a$ we mean the complex conjugate of $a$ so that, for example, ${\bar E}_{n,m} =  e^{-i(kn-\omega(k) m)} = (E_{n,m})^{-1}$, and the positive numbers $\beta_s$ are such that $\beta_0 = 1$ and $\beta_j = j$. The discrete slow varying variables $n_1$, $m_1$ and $m_2$ are defined in terms of $n$ and $m$ by
\bea \nonumber
n &=& n_1 \frac{N^2}{M_1}, \\ \label{3.10a}
m &=& m_1 \frac{N^2}{M_2}, \\ \nonumber
m &=& m_2 N^4.
\eea
Eq. (\ref{3.10a}) is meaningful if $M_1$ and $M_2$ are divisors of $N^2$.

Introducing the expansion (\ref{3.9a}) into eq. (\ref{3.2}) and picking out the coefficients of the various harmonics $(E_{n,m})^s$ we get a set of determining equations. For $s=1$, having defined $\psi^{(1)}= \psi$, we get at lowest order in $\varepsilon$:
\bea \label{3.10}
\psi_{n_1,m_1,m_2} {\Bigl [} (q-p) (1 - z \Omega) - (p + q) ( \Omega - z) {\Bigr ]} = 0, 
\eea
which is identically solved by the dispersion relation (\ref{3.9}). At $\varepsilon^2$ we get a linear equation 
\bea \label{3.11}
M_2 {\Bigl [} (p-q)   z \Omega - (p+q)  \Omega ) {\Bigr ]} {\Bigl [}  \psi_{n_1,m_1+1,m_2} - \psi_{n_1,m_1-1,m_2}  {\Bigr ]}  + \\ \nonumber + M_1 {\Bigl [} (p-q)  z \Omega  + (p+q) z {\Bigr ]} {\Bigl [}  \psi_{n_1+1,m_1,m_2} - \psi_{n_1-1,m_1,m_2} {\Bigr ]} = 0
\eea 
whose solution is given by choosing
\bea \label{3.12}
\psi_{n_1,m_1,m_2} = \phi_{n_2,m_2}
\eea
 where 
 \bea \label{3.13}
 n_2 = n_1 - m_1.
 \eea
The solution (\ref{3.12}) is obtained by choosing  the integers $M_1$ and $M_2$ as
\bea \label{3.14}
M_1 = S ~ \Omega {\Bigl [} (p-q) z  - (p+q) {\Bigr ]}, \\ \nonumber
M_2 = S~ z {\Bigl [} (p-q) \Omega  + (p+q) {\Bigr ]},
\eea
 where $S$ is an arbitrary complex constant. Let us notice that also $n_2 = n_1 + m_1$ solves eq. (\ref{3.13}) by an appropriate choice of  $M_1$ and $M_2$.  Moreover, $\frac{M_1}{M_2} = \omega_{,k}$, the group velocity. As $M_1$ and $M_2$ are integers, not all values of $k$ are admissible as $\omega_{,k}$ must be a rational number.
 
 At $\varepsilon^3$ we get a nonlinear equation for $\phi_{n_2,m_2}$ which depends on $\psi^{(2)}_{n_2,m_2}$ and $\psi^{(0)}_{n_2+1,m_2} - \psi^{(0)}_{n_2-1,m_2}$:
 \bea \label{3.15a}
&& c_1 ( \phi_{n_2,m_2+1} - \phi_{n_2,m_2} ) +  c_2 ( \phi_{n_2+2,m_2} + \phi_{n_2-2,m_2} - 2 \phi_{n_2,m_2} )  + \\ \nonumber &&+ c_3 ( \phi_{n_2+1,m_2} + 
 \phi_{n_2-1,m_2} - 2 \phi_{n_2,m_2} ) +c_4 \phi_{n_2,m_2} 
  (\psi^{(0)}_{n_2+1,m_2} - \psi^{(0)}_{n_2-1,m_2}) + \\ \nonumber &&+ c_5  \psi^{(2)}_{n_2,m_2} {\bar \phi}_{n_2,m_2} = 0,
\eea 
 where
 \bea \nonumber
 c_1 &=&  [(p-q)-z(p+q)], \\ \nonumber
 c_2 &=& S^2 z^2 p q (p-q) {\Bigl [} \frac{(p+q)z - (p-q)}{(p+q) - z(p-q)} {\Bigr ]}^2, \\ \nonumber
 c_3 &=& - 2 S^2 z p q (p-q) \frac{[z(p+q)-(p-q)][(z^2+1)(p+q)-2(p-q)]}{[(p+q)-z(p-q)]^2}, \\ \nonumber
 c_4 &=& - 2 S^2 (p^2-q^2) (z^2-1)   {\Bigl [} \frac{z +1}{(p+q) - z(p-q)} {\Bigr ]}^2, \\ \nonumber
 c_5 &=& -2 \frac{q (p^2-q^2)}{z [(p+q)z-(p-q)]}  {\Bigl [} \frac{(z +1)(z^2-1)}{(p+q) - z(p-q)} {\Bigr ]}^2.
 \eea
 The lowest order equations for the harmonics $s=0$ and $s=2$ appear at $\varepsilon^2$ and give:
  \bea \label{3.15b}
&&  \psi^{(0)}_{n_2+1,m_2} - \psi^{(0)}_{n_2-1,m_2} = 2 |\phi_{n_2,m_2}|^2 \frac{(1+z)^2}{S p z [ (p+q)z -(p-q)]}, \\ \label{3.15c}
&& \psi^{(2)}_{n_2,m_2}  = (\phi_{n_2,m_2})^2 \frac{1+z}{2 p (1-z)}.  
\eea  
Taking these results into account the nonlinear equation (\ref{3.15a}) for $\phi_{n_2,m_2}$ reads:
 \bea \label{3.15}
 i  {\Bigl [} \phi_{n_2,m_2+1} - \phi_{n_2,m_2}  {\Bigr ]} = C_1(k) {\Bigl [} \phi_{n_2+2,m_2} + \phi_{n_2-2,m_2} - 2 \phi_{n_2,m_2}  {\Bigr ]} + \\ \nonumber + C_2(k) {\Bigl [}  \phi_{n_2+1,m_2} + \phi_{n_2-1,m_2} - 2 \phi_{n_2,m_2} {\Bigr ]} + C_3(k) ~\phi_{n_2,m_2} |\phi_{n_2,m_2}|^2,
 \eea 
where  $C_3(k)$ is a real coefficient given by
\bea \label{3.16}
C_3(k) = 2\,{\frac {\sin \left( k \right)  \left( 1+\cos \left( k \right) 
 \right) ^{2}q\, \left( {p}^{2}-{q}^{2} \right) }{p\, \left( ({p}^{2} - q^2)^2
 \left( \cos \left( k \right)  \right) ^{2}-2\,({p}^{4} - q^4) \cos \left( k \right) +({p}^{2}+{q}^{2})^{2}\right) }}.
\eea
The coefficients $C_1(k)$ and $C_2(k)$ are complex and depend on $S$. They are:
\bea \label{3.17}
C_1(k) = i \,q\,p\,{z}^{2} {S}^{2} \left( p-q \right) {\frac { \left(z (p+q)-(p-q) \right)  }{ \left( (p-q)z-(p+q) \right) ^{2}}},
 \eea
 \bea \label{3.18}
 C_2(k) = -2\,i\,q\,p\,z{S}^{2}  \left( p-q \right) {\frac { \left( (p+q)(1+z^2) -2\,(p-q)\,z \right) }{ \left( (p-q)z-(p+q)\right) ^{2}}}.
\eea
Eq. (\ref{3.15}) is a {\it completely discrete  and local} NLS equation depending on 
the first and second  neighboring lattice points. At difference from the Ablowitz and Ladik \cite{al} discrete NLS, the nonlinear term is completely local.
\section{Discussion of the results and conclusive remarks} \label{conclusions}
The choice of the order of slowlyness is essential in defining the points involved in the resulting equation. In our calculation of the multiple-scale reduction of the lpKdV we choose to use the minimum number of points in the various lattices introduced. Here we started from just four points and got a scheme which involves six points. Moreover while the starting initial problem is defined on a staircase,  eq. (\ref{3.15}) is defined on a line. By choosing slow varying functions of higher order, essential for example to get the higher order terms in the expansion necessary to go beyond the NLS \cite{dms} even at the lowest order we would get a nonlinear difference equation involving many more lattice points. This seems to be a peculiarity of the multiple-scale expansion on the lattice. 

This work open a research field of great interest both for the possible mathematical results and for the physical applications. To show this one presents in the following a detailed list of open problems and remarks on which work is in progress:
\begin{enumerate}
\item Prove the integrability of eq. (\ref{3.15}) by reducing the Lax pair of the lpKdV or by constructing its generalized symmetries.
\item Eq. (\ref{3.15}) is invariant with respect to time translation. One can thus reduce it with respect to this Lie point symmetry and get:
\bea \label{c2}
 {\Bigl [} \phi_{n_2+2} + \phi_{n_2-2} - 2 \phi_{n_2}  {\Bigr ]}  &+& d_1 {\Bigl [}  \phi_{n_2+1} + \phi_{n_2-1} - 2 \phi_{n_2} {\Bigr ]} + \\ \nonumber &+& d_2 ~\phi_{n_2} |\phi_{n_2}|^2 = 0.
 \eea 
One would like to show that  equation (\ref{c2}) possesses the Painlev\'e property. On the lattice this is given by the {\it singularity confinement} \cite{ramani,ramani1}  or {\it algebraic entropy} \cite{viallet}.
\item Eq. (\ref{3.15}) has a natural semi-continuous limit when $m_2 \rightarrow \infty$ as $H_2 \rightarrow 0$ in such a way that $t_2 = m_2 H_2$ is finite. In such a case eq. (\ref{3.15}) reduces to the nonlinear differential difference equation 
\bea \label{c1}
i \frac{d \phi_{n_2}}{d{t_2}} &=& e_1 {\Bigl [} \phi_{n_2+2} + \phi_{n_2-2} - 2 \phi_{n_2}  {\Bigr ]} + \\ \nonumber &+& e_2 {\Bigl [}  \phi_{n_2+1} + \phi_{n_2-1} - 2 \phi_{n_2} {\Bigr ]} + e_3 ~\phi_{n_2} |\phi_{n_2}|^2
\eea
If eq. (\ref{3.15}) is integrable than eq. (\ref{c1}) should be an integrable multiple-scale reduction for differential difference equations \cite{leon,lh,yamilov,sakovich} like the Toda lattice.
\item Do the multiple-scale reduction of other integrable lattice equations like the time discrete Toda lattice, the lattice mKdV or the discrete sine--Gordon equation and see what one gets. One could  obtain  eq. (\ref{3.15}) but, maybe some other integrable lattice NLS like equation can be obtained.
\item Do the multiple-scale reduction of the discrete Burgers equation
\bea \label{4.1}
b_{n,m+1} = \frac{b_{n-1,m} [ 1 + b_{n,m} b_{n+1,m} + \alpha b_{n,m}]}{1 + b_{n,m} b_{n-1,m} + \alpha b_{n-1,m}}
\eea
and get a discrete Eckhaus equation \cite{calogero}.
\item Apply the reduction technique to some nonintegrable equation of physical interest like for example those obtained in the case of discrete phenomena in liquid crystals \cite{assanto} and obtain approximate theoretical solutions of the physical result.
\end{enumerate}
\section*{Acknowledgments}

The research presented here benefitted from the NATO collaborative grant
PST.CLG. 978431
 and  was partially supported by  PRIN Project "SINTESI-2004"  of the  Italian Minister for  Education and Scientific Research and from  the Projects {\sl Sistemi dinamici nonlineari discreti: 
simmetrie ed integrabilit\'a} and {\it Simmetria e riduzione di equazioni differenziali di interesse fisico-matematico} of GNFM--INdAM. The author acknowledges fruitful discussions with F. Calogero, R. Hernandez Heredero, J. Hietarinta, O. Ragnisco, M.A. Rodriguez and P.M. Santini.

\end{document}